\documentclass[12pt,a4paper]{article}
\usepackage{jheppub}
\usepackage{dsfont}
\usepackage{amsmath,amssymb,amscd,amsfonts}

\usepackage[toc,page]{appendix}
\usepackage{epsfig}
\usepackage{epstopdf}
\usepackage{latexsym}
\usepackage{graphicx}
\usepackage{subfigure}
\usepackage{booktabs}
\usepackage{bbm}
\usepackage{color}
\usepackage{physics}
\usepackage{tikz}
\usetikzlibrary{matrix}
\usetikzlibrary{decorations.markings,calc,shapes,decorations.pathmorphing,patterns}
\usetikzlibrary{positioning}

\makeatletter
\def\@fpheader{\relax}
\makeatother

\subheader{\begin{flushright}
UTTG-02-20\\
\end{flushright}}

\title{Quotient-AdS/BCFT: Holographic Boundary CFT$_2$ on AdS$_3$ Quotients}
\author{Sanjit Shashi}
\affiliation{Theory Group, Department of Physics, University of Texas, Austin, TX 78712, USA.}

\emailAdd{sshashi[at]utexas.edu}

\abstract{The holographic principle, being a generic feature of quantum gravity, should allow for the consideration of dualities other than AdS/CFT. The AdS/BCFT correspondence, in which the dual field theory has local conformal symmetry and is defined on a manifold with boundary, is one such example. Inspired by the quotienting of AdS$_3$ by spacetime isometries in order to construct multiboundary wormholes dual to multipartite CFT$_2$ states, we find that this correspondence can be understood by combining AdS/CFT with some appropriate quotient procedure. Furthermore, in three bulk dimensions, we find example quotient spaces of AdS$_3$ in order to construct ``natural" bulk duals for specific BCFT$_2$ vacuum states, one of which appears to describe a novel, time-dependent BCFT$_2$ solution. We call this particular refinement of AdS/BCFT, in which we use quotients, the quotient-AdS/BCFT correspondence.}

\begin{document}	
\maketitle
\flushbottom

\section{Introduction}\label{intro}

The AdS/CFT correspondence is perhaps one of the most well-understood manifestations of the holographic principle. However, holography itself is fundamental aspect of any quantum theory of gravity. In particular, for a theory of quantum gravity defined on some number of dimensions and with some spectrum of fields and operators, there is a \textit{dual} description by a quantum field theory on a lower-dimensional space. Of particular note, there is a \textit{one-to-one correspondence} between the fields and operators of the two theories, with empty AdS corresponding to a vacuum CFT state and other CFT states corresponding to quantum gravity states on an AdS background.

In the context of AdS/CFT specifically, this statement of the holographic principle ends up being much more constrained. First, instead of the theory generically being a $(d+1)$-dimensional gravity theory, it is defined on AdS$_{d+1}$. However, AdS$_{d+1}$ is maximally symmetric, so it appears the same around every point. We thus need to ask, where could the degrees of freedom of a dual field theory live?

This is where the work of Brown and Henneaux in \cite{Brown:1986nw} comes into the picture, essentially establishing the notion of the \textit{bulk} and the \textit{conformal boundary},\footnote{The term ``boundary" will be used to refer to two different things in this work: \textit{``conformal" boundary} and \textit{``topological" boundary}, the latter of which refers to the closure of a space minus its interior. We will typically refer to the former type by ``conformal boundary" and the latter type by either just ``boundary" or, for reasons that will become clear, ``defect."} with the latter being placed at spatial infinity under the \textit{Brown-Henneaux boundary conditions}. Furthermore, the field theory restricted to this conformal boundary has symmetries linked to the isometries of AdS$_{d+1}$. The isometry group of AdS$_{d+1}$, SO$(2,d)$, is the \textit{$d$-dimensional global conformal group}, and the symmetry algebra of the conformal boundary field theory comes from enhancing the Lie algebra so$(2,d)$. Thus, the field theory is a CFT$_d$; the AdS/CFT correspondence goes even further in saying that this field theory is a \textit{dual description} of the bulk theory.

The basic idea of this work is to combine the standard view of AdS$_3$/CFT$_2$ as a bulk/boundary duality under the Brown-Henneaux boundary conditions with the quotienting procedure used to construct multiboundary wormholes in \cite{Banados:1992gq,Skenderis:2009ju,Balasubramanian:2014hda,Caceres:2019giy}, with the ultimate goal being a holographic realization of other field theories. Specifically, we will explore more general quotients of AdS$_3$ which are not a part of the multiboundary story. Note that we are working in pure $2+1$ gravity, so we \textit{only} need to consider spaces which are locally AdS$_3$ because there are no bulk degrees of freedom.

We know from these works that quotienting empty AdS$_3$ in a way that does not produce fixed point singularities can yield smooth, locally/asymptotically AdS$_3$ wormhole geometries. Such spaces themselves are dual to states living in multipartite conformal field theories, with the number of asymptotic AdS$_3$ regions matching the number of factors in the tensor product decomposition of the state's Hilbert space. Perhaps the most well-known and well-studied example is the \textit{two-sided BTZ wormhole}, which is dual to the thermofield double state. This state is an entangled one living in the tensor product of two CFT$_2$ Hilbert spaces. \cite{Balasubramanian:2014hda} discusses the three-boundary scenario, as well; the dual state for such a geometry lives in a tensor product of three CFT$_2$ Hilbert spaces.

Another interpretation of this result is that, if we quotient AdS$_3$, the resulting orbifolded space can still host a quantum gravity theory which is dual to a field theory on the conformal boundary. Quantum gravity and gauge theories on orbifolded spaces are nothing new, having been explored in specific cases by \cite{Martinec:2001cf,Horowitz:2001uh}. With regards to multiboundary wormholes, however, these spaces have asymptotic regions consisting of separate copies of AdS$_3$. Thus, under the inherited Brown-Henneaux boundary conditions, AdS$_3$/CFT$_2$ would imply a duality between a quantum gravity theory on an $n$-boundary wormhole and a field theory whose states come from a tensor product of $n$ CFT$_2$ Hilbert spaces.

But, we can ask what happens if we instead \textit{allow} fixed point singularities in the quotient space, particularly on the conformal boundary. Pictorially, such singularities can be seen to arise in an AdS$_3$ quotient if there is no fundamental domain which can exclude them, so it is reasonable to think that the resulting space will have an entire subspace consisting of these singularities. If that subspace has codimension $1$, both the conformal boundary and the bulk would be expected to have a nonempty \textit{(topological) boundary} serving as a \textit{defect} in the theory. Furthermore, by the usual asymptotic procedure, the field theory should still have some local conformal symmetry away from these defects. All of this indicates that, if our quotient space includes fixed point singularities on a codimension $1$ region, the resulting dual field theory is a \textit{boundary CFT$_2$ (BCFT$_2$)}, a type of field theory introduced in \cite{Cardy:2004hm}.

Holographic constructions of BCFTs are nothing new. There is a well-known correspondence known as \textit{AdS/BCFT}, first discussed in a more stringy context by \cite{Karch:2000gx}, and later described in a topologically constructive way by \cite{Takayanagi:2011zk,Fujita:2011fp}. In the latter work specifically, the holographic dual of a given BCFT is realized by essentially ``gluing" it to an appropriate\footnote{``Appropriate" means that the ``total" boundary of the bulk, including both its conformal and topological boundaries, can be broken into two attached pieces, one of which is the region on which the dual BCFT resides, while the other is an \textit{end-of-the-world (ETW) brane}.} asymptotically AdS space. This correspondence has been explored in multiple dimensions and with various configurations for the boundary of the field theory\cite{Fujita:2011fp}. Furthermore, the correspondence has been tested and explored through the calculations of correlation functions\cite{Alishahiha:2011rg} and entropy\cite{Cavalcanti:2018pta,Sato:2020upl,Sully:2020pza}.

However, quotients provide an alternative perspective on how to construct dual spaces to particular BCFT states. Instead of starting with a BCFT defined on some space and gluing a bulk to it, we use identifications by isometries in order to arrive at a BCFT vacuum state with a \textit{particular} domain and a \textit{particular} bulk to serve as the background of a dual quantum gravity theory. In this sense, the use of quotients allows for a refinement of AdS/BCFT. Additionally, we can go in the reverse direction of \cite{Takayanagi:2011zk}, starting with interesting bulk configurations to obtain corresponding BCFT states.

This work is organized as follows. In Section \ref{boundaryCFT}, we start by discussing how a BCFT may be realized by taking a CFT and quotienting the background space. Next, in Section \ref{adsQuotient}, we discuss specific examples of relevant AdS$_3$ quotients whose conformal boundaries would host a BCFT$_2$. Lastly, in Section \ref{holography}, we tie our results to holography and the AdS/BCFT story, discussing how our construction meshes with Takayanagi's. We also briefly address the more general notion of \textit{defect CFTs (dCFTs)} and their role in this story. To avoid ambiguity, we will denote our approach to holographic BCFTs by \textit{quotient-AdS/BCFT}.

\section{BCFTs as CFTs on Orbifolds}\label{boundaryCFT}

If we have a field theory defined on a background manifold $\mathcal{M}$ imbued with a metric, then we first quotient the manifold by some discrete subgroup of isometries $\Gamma$. On the resulting orbifold $\mathcal{M}/\Gamma$, fields and operators from the original theory which respect the symmetries in $\Gamma$ will be single-valued, while those which do not will have \textit{branches}. So, we can obtain a well-defined theory on $\mathcal{M}/\Gamma$ by discarding the objects which are \textit{not} invariant under $\Gamma$.\footnote{Two different field theories may, upon quotienting by isometries, give rise to the same field theory on homeomorphic orbifolds.}

Consider a simple example, taking two-dimensional Euclidean space, $\mathbb{R}^2$ imbued with the metric,
\begin{equation}
ds_{2}^2 = dX^2 + dY^2.
\end{equation}

We observe that parity in $Y$, $P_Y(X,Y) = (X,-Y)$, is an isometry of $\mathbb{R}^2$. If we consider a field $\Phi$ on $\mathbb{R}^2$ such that, for all points in $\mathbb{R}^2$,
\begin{equation}
\Phi(X,Y) = \Phi(X,-Y),
\end{equation}
then $\Phi$ will be single-valued on $\mathbb{R}^2/P_Y$. However, if there is at least one point $(X_0,Y_0)$ at which,
\begin{equation}
\Phi(X_0,Y_0) \neq \Phi(X_0,-Y_0),
\end{equation}
then $\Phi$ will be multi-valued at this point in $\mathbb{R}^2/P_Y$. The same also holds for operators, so, in order to have a well-defined theory on $\mathbb{R}^2/P_Y$, we must take fields and operators which are invariant under $P_Y$.

In this example, note that the locus of points $(X,0)$ for all $X \in \mathbb{R}$ forms the set of fixed points of $P_Y$. Thus, $\mathbb{R}^2/P_Y$ is actually the Euclidean upper half-plane whose interior is locally $\mathbb{R}^2$, and theories on the Euclidean upper half-plane are obtained from theories on $\mathbb{R}^2$ and the operator $P_Y$. This orbifold can be realized via the ``folding" discussed in \cite{Bachas:2001vj}.

For metric spaces, as isometries locally leave the metric invariant, quotienting by a discrete subgroup of isometries ensures that the orbifold $\mathcal{M}/\Gamma$ has the same local structure as $\mathcal{M}$ everywhere except for at defects. In fact, for maximally symmetric $\mathcal{M}$, neighborhoods of non-defect points in $\mathcal{M}/\Gamma$ are themselves homeomorphic to $\mathcal{M}$. This is certainly the case for $\mathbb{R}^2/P_Y$, and it is also the case for such quotients of AdS$_3$. Indeed, this why the multiboundary wormholes constructed as quotients of AdS$_3$ are also locally AdS$_3$.\footnote{See \cite{Banados:1992gq,Skenderis:2009ju,Balasubramanian:2014hda,Caceres:2019giy} for details.}

Thus, if the original theory has more (local) symmetry, then a field theory on an orbifold will inherit that symmetry, as well. A simple example explored in this work will be a BCFT$_2$ on the upper half-plane. Such a theory consists of states which are both well-defined on $\mathbb{R}^2/P_Y$ and which have local conformal symmetry. As such, we could think about a BCFT$_2$ as being obtained by taking a CFT$_2$ on the plane and quotienting by $P_Y$.

With that, we have arrived at the main purpose of this work. We have a procedure by which we can construct BCFTs from CFTs, and we have the AdS/CFT correspondence. Combining the two, we should be able to realize holographic duals of BCFTs states as quotient spaces of AdS. We will specifically focus on the case of AdS$_3$, but the above arguments are much more general.

\section{AdS$_3$ Quotients with Fixed Points}\label{adsQuotient}

We now study quotient spaces of AdS$_3$ which have fixed points, both in the bulk and on the conformal boundary. Specifically, we will quotient AdS$_3$ by elements of its \textit{full} SO$(2,2)$ isometry group; such quotient spaces will be locally AdS$_3$ everywhere \textit{except for} at any defects, such as fixed points, which will comprise ETW branes.

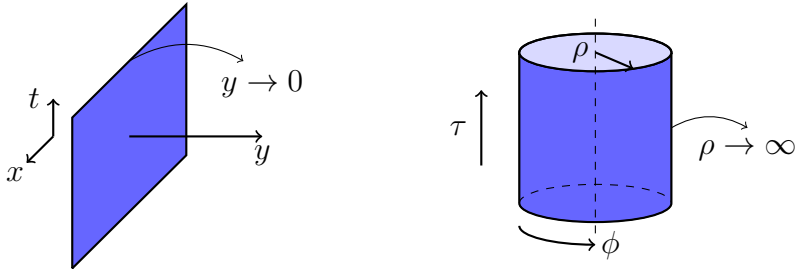
\begin{figure*}
\centering
\begin{tikzpicture}
\draw[->,thick] (-1,1) to (-1,1.5);
\draw[->,thick] (-1,1) to (-1.35,0.65);
\draw[-,thick,fill=blue!60] (-0.75,-0.75) to (0.75,0.75) to (0.75,2.75) to (-0.75,1.25) to (-0.75,-0.75);

\draw[->,thick] (0,0+1) to (1.75,0+1);
\node at (-1.5,0.5) {$x$};
\node at (-1.25,1.5) {$t$};
\node at (1.75,0.75) {$y$};

\draw[->] (0,2) to[bend left] (1.5,2);
\node at (1.75,1.7) {$y \to 0$};
\end{tikzpicture}\qquad\qquad
\begin{tikzpicture}
\draw[-,thick,fill=blue!60] (-1,0) arc (180:360:1 and 0.25) to (1,2) arc (360:180:1 and 0.25) to (-1,0);
\draw[-,thick,fill=blue!15] (0,2) ellipse (1 and 0.25);
\draw[-,thick] (1,2) arc (0:180:1 and 0.25);

\draw[-,dashed] (-1,0) arc (180:0:1 and 0.25);
\draw[-] (-1,0) arc (180:360:1 and 0.25);
\draw[-] (-1,0) to (-1,2);
\draw[-] (1,0) to (1,2);
\draw[-] (0,2) ellipse (1 and 0.25);

\draw[->,thick] (-1.5,0.5) to (-1.5,1.5);
\node at (-1.8,1) {$\tau$};

\draw[->,thick] (-1,-0.3) arc (180:270:1 and 0.25);
\node at (0.2,-0.55) {$\phi$};

\draw[->,thick] (0,2) to (0.5,2-0.216);
\node at (-0.2,2) {$\rho$};

\draw[->] (1,1) to[bend left] (2,1);
\node at (2,0.7) {$\rho \to \infty$};

\draw[-,dashed] (0,-0.4) to (0,2.5);
\end{tikzpicture}
\caption{On the left, we have AdS$_3$ depicted in terms of Poincar\'e coordinates. On the right, we have AdS$_3$ depicted in global coordinates. The conformal boundaries are in blue; they are at $y \to 0$ and $\rho \to \infty$, respectively.}
\label{figs:adscoords}
\end{figure*}

Before delving into explicit constructions, however, we review some basic details of AdS$_3$. There are two coordinate systems which we will use: \textit{Poincar\'e coordinates},
\begin{equation}
\frac{ds^2}{\ell^2} = \frac{-dt^2 + dx^2 + dy^2}{y^2},\label{poincare}
\end{equation}
in which $t,x \in \mathbb{R}$ and $y > 0$, and \textit{global coordinates},
\begin{equation}
\frac{ds^2}{\ell^2} = -\cosh^2\rho d\tau^2 + d\rho^2 + \sinh^2\rho d\phi^2,\label{global}
\end{equation}
in which $\tau \in \mathbb{R}$, $\rho \geq 0$, and $\phi \sim \phi + 2\pi$. We visually represent both coordinate systems in Figure \ref{figs:adscoords}.

With Poincar\'e coordinates, we will primarily discuss the isometries used to quotient AdS$_3$; particularly, we will use its natural foliation of AdS$_3$ into copies of the \textit{Poincar\'e upper half-plane}, i.e. the hyperbolic plane $\mathbb{H}$, in order to better visualize the actions of the isometries. We can also use Poincar\'e coordinates to understand what the bulk spacetime and conformal boundary both look like after quotienting. To do so, note that the conformal boundary is at $y = 0$, at which \eqref{poincare} becomes,
\begin{equation}
\frac{ds^2}{\ell^2} \xrightarrow{y \to 0} \frac{1}{y^2}(-dt^2 + dx^2).\label{confboundP}
\end{equation}

\eqref{confboundP} is conformally equivalent to a two-dimensional flat spacetime. By analytically continuing the Poincar\'e $t$ coordinate to imaginary time, we confirm that the dual state is the vacuum of a CFT$_2$ on $\mathbb{R}^2$. Furthermore, we can relate the identifications performed in Poincar\'e coordinates to identifications in coordinates of this plane.

However, in global coordinates, the conformal boundary is compactified along one of its dimensions, allowing us to visualize the entire space more easily and providing a more satisfying picture for the quotiented bulk. Additionally, Poincar\'e coordinates only cover part of the full spacetime. To switch from Poincar\'e coordinates to global coordinates, we use,\footnote{The relationship between Poincar\'e and global coordinates can be found by the standard treatment of AdS$_3$ as a hyperboloid embedded in an ambient $(2+2)$-dimensional flat spacetime. As done in \cite{Banados:1992gq}, the ambient coordinates can be written as functions of the global coordinates, while the Poincar\'e coordinates can be written as functions of the ambient coordinates.}
\begin{align}
t &= -\frac{\ell\cosh\rho\cos\tau}{\cosh\rho\sin\tau+\sinh\rho\cos\phi},\label{transPG1}\\
x &= \frac{\ell\sinh\rho\sin\phi}{\cosh\rho\sin\tau+\sinh\rho\cos\phi},\label{transPG2}\\
y &= \frac{\ell}{\cosh\rho\sin\tau+\sinh\rho\cos\phi}.\label{transPG3}
\end{align}

Furthermore, in global coordinates, we require the conformal transformation between the cylindrical conformal boundary shown in Figure \ref{figs:adscoords} and the Euclidean plane. First, taking $\rho \to \infty$, \eqref{global} becomes,
\begin{equation}
\frac{ds^2}{\ell^2} \xrightarrow{\rho \to \infty} \frac{e^{2\rho}}{4}(-d\tau^2 + d\phi^2).\label{confbound1}
\end{equation}

\eqref{confbound1} is conformally equivalent to a cylinder with a timelike coordinate. So, we now Euclideanize $\tau$ as,
\begin{equation}
\tau \to -i\tau_E,\label{wick}
\end{equation}
and define coordinates $(x_E,y_E) \in \mathbb{R}^2$ by,
\begin{align}
&x_E + iy_E = e^{\tau_E+i\phi}\nonumber\\
&\implies x_E = e^{\tau_E}\cos\phi,\ \ y_E = e^{\tau_E}\sin\phi.\label{cyltoplane}
\end{align}

Observe that,
\begin{align}
dx_E^2 + dy_E^2
&= (x_E^2 + y_E^2)(d\tau_E^2 + d\phi^2)\nonumber\\
&= e^{2i\tau}(-d\tau^2 + d\phi^2).\label{confbound2}
\end{align}

Thus, we have that the $(x_E,y_E)$-plane spanning $\mathbb{R}^2$ is conformally equivalent to the Euclidean cylinder; analytically continuing back to Lorentzian time, we then deduce that the plane is conformally equivalent (up to analytic continuation) to the boundary of AdS$_3$ in \eqref{confbound1}. The relationship between the coordinates of the plane and the coordinates of AdS$_3$ is expressed in \eqref{cyltoplane}, allowing us to relate any identifications performed in global coordinates directly to identifications performed on the plane.

Lastly, in both of the following constructions, we will use the \textit{parity operator}. As such, these isometries cannot be obtained from a Killing vector. This is a major departure from other work which has been done on quotient spaces of AdS$_3$ in \cite{Banados:1992gq,Skenderis:2009ju,Balasubramanian:2014hda,Caceres:2019giy}, all of which discuss how multiboundary wormhole configurations (which do not have any fixed point singularities) arise from quotients by Killing vectors, i.e. by isometries in the \textit{identity component} of SO$(2,2)$.

Note that a class of quotient spaces exhibiting axisymmetry and involving parity have been explored in the past\cite{Loran:2010qn,Loran:2010zy}. Those constructions have fixed points which live on one of four types of surfaces: a spacelike plane, a spacelike cylinder, a null plane, or a null cylinder. Our constructions differ in that the fixed-point surfaces we find are timelike, but the key commonality is that quotienting by parity leads to fixed points.

\subsection{AdS$_3$ Half-Space from Parity}\label{parityQ}

We start by quotienting by parity, which we find will produce the $\rho_* = 0$ \textit{AdS$_3$ half-space} configuration discussed in \cite{Takayanagi:2011zk,Fujita:2011fp}. In Poincar\'e coordinates, this is,
\begin{equation}
(t,x,y) \to (t,-x,y).\label{parity}
\end{equation}

\begin{figure*}
\centering
\begin{tikzpicture}
\draw[->,thick,blue!60] (-3.5,0) to (3.5,0);
\draw[->,thick,color=red] (0,0) to (0,3.5);
\node[color=red] at (0,0) {$\bullet$};

\draw[->,very thick] (4,1.75) to (6,1.75);

\draw[->,thick,blue!60] (7,0) to (10.5,0);
\draw[->,thick,color=red] (7,0) to (7,3.5);
\node[color=red] at (7,0) {$\bullet$};

\node at (-2,2) {$\bullet$};
\node at (2,2) {$\bullet$};
\node at (9,2) {$\bullet$};
\end{tikzpicture}
\caption{A constant $t$ slice of AdS$_3$ on the left, and its quotient by \eqref{parity} on the right. The red represents the fixed points of this isometry, which, upon quotienting, make-up a boundary. The black dots are points which are identified with one another. The blue consists of points on the conformal boundary. This is analogous to the ``folding" in \cite{Bachas:2001vj}.}
\label{figs:parity1}
\end{figure*}
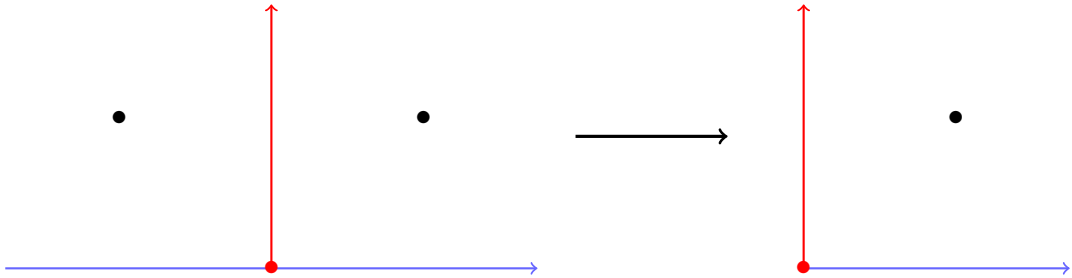

Not only is \eqref{parity} an obvious isometry of \eqref{poincare}, but its action on and fixed points in each constant $t$ slice are also evident. Every point is reflected across the $x$-axis of each copy of $\mathbb{H}$, so the points along the $x$-axis itself are all fixed by this isometry. We show this in Figure \ref{figs:parity1}.

By considering \eqref{confboundP}, we can see that the conformal boundary of the quotient space is the Euclidean upper half-plane. Indeed, in Figure \ref{figs:parity1}, if we ``stack" the folded constant $t$ slices, the resulting spacetime's conformal boundary will just be an upper half-plane, with the fixed points living along an AdS$_2$ ETW brane $x = 0$. We will corroborate this statement in global coordinates. From \eqref{transPG1}-\eqref{transPG3}, parity in global coordinates is realized as,
\begin{equation}
(\tau,\rho,\phi) \to (\tau,\rho,2\pi-\phi).\label{parityG}
\end{equation}

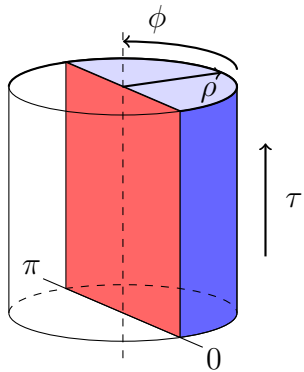
\begin{figure}
\centering
\begin{tikzpicture}[scale=1.5]
\draw[-,fill=blue!60] (0.5,-0.216506) arc (300:360:1 and 0.25) to (1,2) arc (360:300:1 and 0.25) to (0.5,-0.216506);
\draw[-,thick] (1,2) arc (0:180:1 and 0.25);

\draw[-,thick,fill=blue!15] (0.5,2-0.216506) arc (300:300+180:1 and 0.25) to (0.5,2-0.216506);

\draw[-,fill=red!60] (0.5,-0.216506) to (-0.5,0.216506) to (-0.5,2+0.216506) to (0.5,-0.216506+2) to (0.5,-0.216506);

\draw[-,dashed] (-1,0) arc (180:0:1 and 0.25);
\draw[-] (-1,0) arc (180:360:1 and 0.25);
\draw[-] (-1,0) to (-1,2);
\draw[-] (1,0) to (1,2);
\draw[-] (0,2) ellipse (1 and 0.25);

\draw[->,thick] (1.25,0.5) to (1.25,1.5);
\node at (1.5,1) {$\tau$};

\draw[->,thick] (1,2.3-0.15) arc (360:450:1 and 0.25);
\node at (0.3,2.6) {$\phi$};

\draw[->,thick] (0,2) to (0.866025,2+0.125);
\node at (0.75,1.95) {$\rho$};

\draw[-,dashed] (0,-0.4) to (0,2.5);

\draw[-] (0.5+0.2,-0.216506-0.0433012*2) to (-0.5-0.2,0.216506+0.0433012*2);

\node at (0.5+0.2+0.1,-0.216506-0.0433012*2-0.1) {$0$};
\node at (-0.5-0.2-0.1,0.216506+0.0433012*2+0.1) {$\pi$};
\end{tikzpicture}
\caption{The quotient of AdS$_3$ by parity, in global coordinates. The red plane, an AdS$_2$ ETW brane, is the locus of fixed points, and the dark blue portion is the conformal boundary. Precisely half of the points from the bulk remain after quotienting. Note that considering the fixed points as defects in the bulk constructively reproduces the ``slicing" discussed in \cite{Jensen:2013lxa}; the induced metric on the space of fixed points is that of AdS$_2$.}
\label{figs:parity2}
\end{figure}

In global coordinates, the fixed points are all of the points for which $\phi = 0,\pi$. Performing this quotient on the cylindrical visualization of AdS$_3$, we obtain AdS$_3$ half-space as shown in Figure \ref{figs:parity2}.

In order to better understand the conformal boundary of AdS$_3$ half-space, we look at how \eqref{parityG} acts on the $(x_E,y_E)$-plane discussed at the start of the section. Using \eqref{cyltoplane}, we see that,
\begin{equation}
(x_E,y_E) \to (x_E,-y_E).
\end{equation}

Thus, the conformal boundary of half-AdS$_3$ is indeed conformally equivalent to the Euclidean upper half-plane, since we are essentially folding the $(x_E,y_E)$-plane along the $x_E$-axis.

Lastly, we mention that the ETW brane found in this way is similar to those of the Randall-Sundrum model discussed in \cite{Randall:1999vf}. Specifically, we consider an action consisting of an Einstein-Hilbert term, a Gibbons-Hawking-York term, and a codimension-$1$ world-volume, corresponding to a brane. The coupling appearing in the last of these terms is the \textit{tension} $T$, and, in conjunction with the Gibbons-Hawking-York term, one finds that setting a Neumann boundary condition (discussed in \cite{Takayanagi:2011zk}) relates the extrinsic curvature $K_{ab}$ of the brane (defined over the brane's indices) to $T$ and the induced metric $h_{ab}$ by,
\begin{equation}
K_{ab} = (K-T)h_{ab}.
\end{equation}

Upon finding the extrinsic curvature for half-space configurations, \cite{Takayanagi:2011zk} explicitly solves for the brane tension in terms of tune-able parameter $\rho_* \in \mathbb{R}$, which represents the position of the brane.

However, the ETW brane in half-space obtained by quotienting lies at $\rho_* = 0$, for which the tension goes to $0$. In other words, quotienting by parity, an isometry, yields a \textit{tensionless ETW brane} as the locus of fixed points. Classically, our AdS$_3$ half-space configuration is thus a solution to Einstein gravity.

\subsection{AdS$_3$ Strip from Inversion + Parity}\label{parityInvQ}

We can consider a more exotic isometry than just the parity transformation used to construct half-AdS$_3$. First, consider \textit{inversion} in Poincar\'e coordinates,\footnote{We call this transformation such because it reduces to the usual inversion map on $\mathbb{H}$ at $t = 0$.}
\begin{equation}
(t,x,y) \to \frac{a^2}{x^2 + y^2 - t^2}(t,-x,y).\label{inversion}
\end{equation}

Here, $a$ is some real parameter with the same dimensions as the Poincar\'e coordinates. If we further compose \eqref{inversion} with parity, then we obtain,
\begin{equation}
(t,x,y) \to \frac{a^2}{x^2 + y^2 - t^2}(t,x,y).\label{inversionP}
\end{equation}

This map preserves the metric. However, note that it is not a well-defined isometry on all of the Poincar\'e metric in AdS$_3$. In particular, on any constant $t$ slice, if we consider a point $(t,x,y)$ where,
\begin{equation}
x^2 + y^2 - t^2 < 0,\label{exclude}
\end{equation}
then \eqref{inversionP} would map $(t,x,y)$ to a point with a \textit{negative} $y$ coordinate, which is not within the domain of the Poincar\'e coordinates. Thus, we must exclude the points for which \eqref{exclude} holds.

Now, to gain an understanding of where quotienting by \eqref{inversionP} may yield defects, we consider two possible cases for points in AdS$_3$. First, take a point $(t_0,x_0,y_0)$ such that $x_0^2 + y_0^2 - t_0^2 \neq 0$. Then, $(t_0,x_0,y_0)$ is a fixed point of \eqref{inversionP} if and only if,
\begin{equation}
\frac{a^2}{x_0^2 + y_0^2 - t_0^2} = 1 \iff x_0^2 + y_0^2 - t_0^2 = a^2.
\end{equation}

Next, take $(t_0,x_0,y_0)$ such that $x_0^2 + y_0^2 - t_0^2 = 0$. Under \eqref{inversionP}, any such point would go to infinity, so these points are singular. However, because we are \textit{quotienting} by \eqref{inversionP}, these points are also identified with one another. Additionally, as we have removed the points for which \eqref{exclude} holds, the singular points constitute a defect.

To summarize, the locus of fixed points is,
\begin{equation}
\{(t,x,y) \in \text{AdS}_3\,|\,x^2 + y^2 - t^2 = a^2\},\label{locusFP}
\end{equation}
while the locus of singular points is,
\begin{equation}
\{(t,x,y) \in \text{AdS}_3\,|\,x^2 + y^2 - t^2 = 0\}.\label{locusSP}
\end{equation}

By plugging-into \eqref{poincare}, we find that both of these surfaces are AdS$_2$ branes. Specifically, we will show that the fixed points \eqref{locusFP} comprise an ETW brane in the quotient space, but note that we may compute its extrinsic curvature by starting with the normal vector,
\begin{equation}
n_\mu = \frac{1}{y\sqrt{x^2 + y^2 - t^2}}(-t,x,y).
\end{equation}

Then, writing the points on \eqref{locusFP} as $u^a = (t,y)$ and $v^\mu = (t,\pm \sqrt{a^2 + t^2 - y^2},y)$, we write the extrinsic curvature $K_{ab}$ over the brane's indices as,
\begin{equation}
K_{ab} = \pdv{v^\mu}{u^a} \pdv{v^\nu}{u^b} \nabla_\mu n_\nu.
\end{equation}

Plugging-in results in an extrinsic curvature of $0$, so the brane is tensionless, just as in the previously examined case.

Both the fixed points and the singular points live on semicircular geodesics in constant $t$ surfaces. These semicircles are centered at $(0,0)$ and have radii $a^2 + t^2$ and $t^2$, respectively. Thus, both the locus of fixed points and the locus of singular points \textit{shrink} as $t$ goes from $-\infty$ to $0$, then \textit{grow} as $t$ goes from $0$ to $\infty$.

We are ready to describe the quotient space. Essentially, we wish to show that all of the points between the surfaces defined by \eqref{locusFP} and \eqref{locusSP} map to points strictly outside of \eqref{locusFP} in a \textit{one-to-one way}; this would allow us to accurately represent all points in the quotient space. More concretely, we start with the first part of this statement; consider an inner region point $(t_0,x_0,y_0)$ such that,
\begin{equation}
x_0^2 + y_0^2 = R^2,\ \ 0 < R^2 - t_0^2 < a^2.
\end{equation}

Under \eqref{inversionP},
\begin{equation}
(t_0,x_0,y_0) \mapsto \frac{a^2}{R^2 - t_0^2}(t_0,x_0,y_0) = (t_0',x_0',y_0').\label{imageIP}
\end{equation}

So, we want to show that,
\begin{equation}
(x_0')^2 + (y_0')^2 > a^2 + (t_0')^2.\label{ineq}
\end{equation}

By plugging-in \eqref{imageIP} and assuming $R^2 - t_0^2 > 0$, we have that \eqref{ineq} is only true if and only if,
\begin{align}
&\frac{a^4 R^2}{(R^2 - t_0^2)^2} > a^2 + \frac{a^4 t_0^2}{(R^2 - t_0^2)^2}\\
&\iff \frac{a^4 (R^2 - t_0^2)}{(R^2 - t_0^2)^2} = \frac{a^4}{R^2 - t_0^2} > a^2\\
&\iff a^2 > R^2 - t_0^2.\label{conditionTrue}
\end{align}

\eqref{conditionTrue} was our assumption. Thus, on each particular slice, after removing the points for which \eqref{exclude} holds, the fixed and singular points define an ``inner" region and an ``outer" region, defined by,
\begin{align}
\text{Inner:}\ &\{(t,x,y) \in \text{AdS}_3\,|\,0 < x^2 + y^2 - t^2 < a^2\},\\
\text{Outer:}\ &\{(t,x,y) \in \text{AdS}_3\,|\,x^2 + y^2 - t^2 > a^2\},
\end{align}
and the inner region is mapped to the outer region by \eqref{inversionP}.

Now, we show that this mapping is one-to-one; if we have inner region points $(t_1,x_1,y_1)$ and $(t_2,x_2,y_2)$ such that their images are the same, i.e.,
\begin{equation}
\frac{a^2}{x_1^2 + y_1^2 - t_1^2}(t_1,x_1,y_1) = \frac{a^2}{x_2^2 + y_2^2 - t_2^2}(t_2,x_2,y_2),\label{imageInner}
\end{equation}
then, by squaring and summing the components, we may write,
\begin{align}
&\frac{a^4(x_1^2 + y_1^2 - t_1^2)}{(x_1^2 + y_1^2 - t_1^2)^2} = \frac{a^4(x_2^2 + y_2^2 - t_2^2)}{(x_2^2 + y_2^2 - t_2^2)^2}\nonumber\\
&\implies x_1^2 + y_1^2 - t_1^2 = x_2^2 + y_2^2 - t_2^2.\label{sameImg}
\end{align}

However, combining \eqref{sameImg} with \eqref{imageInner} implies that,
\begin{equation}
(t_1,x_1,y_1) = (t_2,x_2,y_2),
\end{equation}
so, if we have two different inner region points, then they must map to two different outer region points.

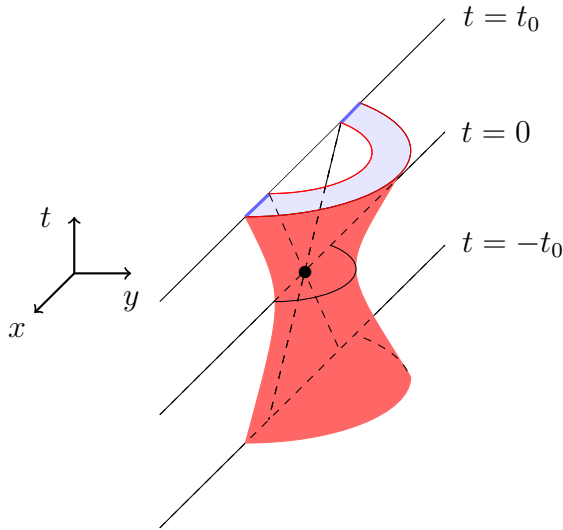
\begin{figure}
\centering
\begin{tikzpicture}[scale=1.5]
\draw[->,thick] (-1-1,1) to (-1-1,1.5);
\draw[->,thick] (-1-1,1) to (-1.5,1);
\draw[->,thick] (-1-1,1) to (-1.35-1,0.65);

\node at (-1.5-1,0.5) {$x$};
\node at (-1.25-1,1.5) {$t$};
\node at (-1.5,1-0.25) {$y$};

\draw[-] (-1-0.25,-1-0.25) to (1+0.25,1+0.25);
\draw[-] (-1-0.25,-0.25) to (1+0.25,2+0.25);
\draw[-] (-1-0.25,1-0.25) to (1+0.25,3+0.25);

\node at (1+0.85,1+0.25) {$t = -t_0$};
\node at (1+0.7,2+0.25) {$t = 0$};
\node at (1+0.75,3+0.25) {$t = t_0$};

\draw[-,fill=red!60,color=red!60] (-0.5,2-0.5) arc (-90:-22:1.45 and 0.585) .. controls (0.6-0.275,1+0.085/2-0.05) and (0.6-0.275,1+0.085/2+0.05) .. (0.95,0.085) arc (0:-90:1.45 and 0.585) .. controls (0-0.15,0.9-0.1) .. (-0.5,2-0.5);

\draw[-,fill=blue!10] (-0.5,2-0.5) arc (-90:46:1.45 and 0.585) to (-0.5,2-0.5);

\draw[-,color=blue!60,very thick] (-0.5,2-0.5) to (0-0.125+0.26125-0.425-0.005,1-0.15+0.26125+0.5875);
\draw[-,color=blue!60,very thick] (0-0.125+0.26125-0.425+0.625-0.005,1-0.15+0.26125+0.5875+0.625) to (1.01-0.5,1.01+2-0.5);

\draw[-,red,fill=white] (-0.5+0.205,2-0.5+0.205) arc (-90:46:1.45/1.6 and 0.585/1.6);

\node at (0-0.125+0.15,1-0.15+0.15) {$\bullet$};

\draw[-] (-0.125-0.07-0.05,0.85-0.05-0.05) arc (-90:46.5:1.45/2.025 and 0.585/2.025);

\draw[-,dashed] (-1-0.25,-0.25) to (1+0.25,2+0.25);
\draw[-,dashed] (-1-0.25,-1-0.25) to (1+0.25,1+0.25);
\draw[-,dashed] (0.95-0.03,0.085+0.05) arc (11:48:1.45 and 0.585);

\draw[-,dashed] (0-0.125+0.26125-0.425,1-0.15+0.26125+0.5875) to (-0.5+0.825,-0.5+0.825);

\draw[-,dashed] (0-0.125+0.26125-0.425+0.625,1-0.15+0.26125+0.5875+0.625) to (-0.5+0.205,-0.5+0.205);

\draw[-,dashed] (0-0.125+0.26125-0.425+0.625,1-0.15+0.26125+0.5875+0.625) to (-0.5+0.205,-0.5+0.205);

\draw[-] (0.33625,2.32375) to (0.33625-0.14,2.32375-0.14*4.14851485149);

\draw[-,color=red] (-0.5,2-0.5) arc (-90:46:1.45 and 0.585);
\draw[-,color=red] (-0.5+0.205,2-0.5+0.205) arc (-90:46:1.45/1.6 and 0.585/1.6);
\end{tikzpicture}
\caption{The quotient of AdS$_3$ by inversion composed with parity, with a cross-section taken at the $t = t_0$ slice. The outer (red) surface is the locus of fixed points \eqref{locusFP} and, consequently, the tensionless ETW brane, while the inner surface is the locus of singular points \eqref{locusSP}, representing infinity. The bulk (depicted in light blue at $t = t_0$) is strictly between the two branes and ``widens" away from $t = 0$. The singular points develop from the bullet point at $t = 0$. The shape is hollow because we remove the points for which \eqref{exclude} holds.}
\label{figs:hyperSurf}
\end{figure}

With that proven, we conclude that the inner and outer regions are identified when quotienting by \eqref{inversionP} in a one-to-one way, allowing us to use just the inner region\footnote{This is a matter of convention; we may also select the outer region.} in order to depict the quotient space. In other words, the inner region, which includes an ETW AdS$_2$ brane along the fixed points, \textit{is} the bulk of the quotient space. This bulk is depicted in Figure \ref{figs:hyperSurf}; we call the geometry an \textit{AdS$_3$ strip}.

AdS$_3$ strips are still covered by the Poincar\'e metric except for at the fixed and singular points, so we can consider what happens along the conformal boundary $y \to 0$. In particular, \eqref{locusFP} describes a hyperbola,
\begin{equation}
\{(t,x,0) \in \text{AdS}_3\,|\,x^2 - t^2 = a^2\}.\label{locusFPB}
\end{equation}

Meanwhile, \eqref{locusSP} describes two lines,
\begin{equation}
\{(t,x,0) \in \text{AdS}_3\,|\,x^2 = t^2\}.\label{locusSPB}
\end{equation}

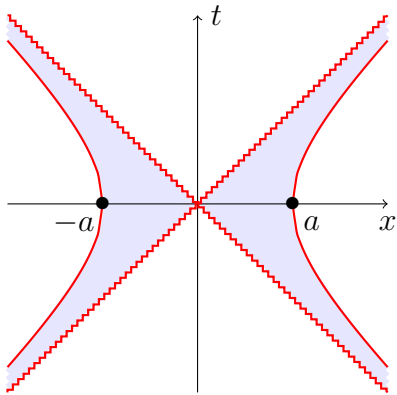
\begin{figure}
\centering
\begin{tikzpicture}[scale=1.25]
\fill[blue!10,decoration = {zigzag,segment length = 1mm, amplitude = 0.25mm},decorate] (0,0) -- (2,2) to (2,1.73205) to (2-1.73205,0) to (0,0);
\fill[blue!10,decoration = {zigzag,segment length = 1mm, amplitude = 0.25mm},decorate] (0,0) -- (2,-2) to (2,-1.73205) to (2-1.73205,0) to (0,0);
\fill[blue!10,decoration = {zigzag,segment length = 1mm, amplitude = 0.25mm},decorate] (0,0) -- (-2,2) to (-2,1.73205) to (-2+1.73205,0) to (0,0);
\fill[blue!10,decoration = {zigzag,segment length = 1mm, amplitude = 0.25mm},decorate] (0,0) -- (-2,-2) to (-2,-1.73205) to (-2+1.73205,0) to (0,0);

\fill[blue!10] (2,1.75) .. controls (0.65,0) .. (2,-1.75) to (0.2,0) to (2,1.75);
\fill[blue!10] (-2,1.75) .. controls (-0.65,0) .. (-2,-1.75) to (-0.2,0) to (-2,1.75);

\draw[->] (0,-2) to (0,2);
\draw[->] (-2,0) to (2,0);

\draw[thick,decoration = {zigzag,segment length = 1mm, amplitude = 0.25mm},decorate,color=red] (-2,-2)--(2,2);
\draw[thick,decoration = {zigzag,segment length = 1mm, amplitude = 0.25mm},decorate,color=red] (-2,2)--(2,-2);

\draw[thick,-,color=red,domain = 1:2,smooth,variable = \x] plot ({\x},{sqrt(\x*\x-1)});
\draw[thick,-,color=red,domain = 1:2,smooth,variable = \x] plot ({\x},{-sqrt(\x*\x-1)});

\draw[thick,-,color=red,domain = -1:-2,smooth,variable = \x] plot ({\x},{sqrt(\x*\x-1)});
\draw[thick,-,color=red,domain = -1:-2,smooth,variable = \x] plot ({\x},{-sqrt(\x*\x-1)});

\node at (1,0) {$\bullet$};
\node at (1.2,-0.2) {$a$};
\node at (-1,0) {$\bullet$};
\node at (-1.3,-0.2) {$-a$};

\node at (2,-0.2) {$x$};
\node at (0.2,2) {$t$};
\end{tikzpicture}
\caption{The dual BCFT$_2$ state obtained upon quotienting by \eqref{inversionP}. The red lines represent defects, with the solid lines being the locus of fixed points \eqref{locusFPB} and the jagged lines being the locus of singular points \eqref{locusSPB}. The fundamental domain is in blue. The parameter $a$ in \eqref{inversionP} controls the minimum separation of the fixed points.}
\label{figs:parityBCFT}
\end{figure}

Thus, the configuration of the dual BCFT$_2$ state is shown in Figure \ref{figs:parityBCFT}. This particular state has time-dependent boundary dynamics; we can interpret the fixed points as Rindler observers and the singular points as setting a ``speed limit" on those observers. Furthermore, we should be able to probe its dynamics by computing quantum information quantities, i.e. complexity or entropy, which we could expect to evolve in time. We leave this to future work.

The AdS$_3$ strip is also interesting from a holographic perspective. It is essentially a hyperbolic shell which ``widens" and ``thins" in time. As such, it should have characteristic geometric quantities that can be computed using the Poincar\'e metric. Upon probing the dual BCFT$_2$ state, we can check how field theoretic quantities may compare to the bulk geometric ones. For instance, setting some cutoff $\delta \ll 1$ so that we consider $y \in [\delta,\infty)$ and taking some (positive) time $t_0 > \delta$, one may use \eqref{poincare} to explicitly compute the area of the inner region at $t = t_0$,
\begin{equation}
a(t_0,\delta) = \frac{2}{\delta}\left(\sqrt{a^2 + t_0^2} - t_0\right) + O(\delta).\label{areaStrip}
\end{equation}

At late times $t_0 \to \infty$, the divergent piece of \eqref{areaStrip} vanishes. We interpret this as a dynamical statement; the degrees of freedom of the boundary state vanish as $t_0 \to \infty$ (and, by symmetry, as $t_0 \to -\infty$).

\section{Holographic Considerations}\label{holography}

Having discussed a couple of AdS$_3$ quotients and their corresponding conformal boundaries, we are ready to discuss duality. Much of this conversation will be kept general, to any number of dimensions.

\subsection{Quotient-AdS$_3$/BCFT$_2$ and AdS/BCFT}\label{qAdSBCFT}

When considering quantum gravity\footnote{The classical bulk theory is taken to be \textit{pure} Einstein gravity, and we consider the quantum gravity theory in the semiclassical limit using the Einstein-Hilbert action.} on AdS$_{d+1}$, AdS/CFT asserts duality with a CFT$_d$ on the conformal boundary, $\mathbb{R}^d$. Quotienting AdS$_{d+1}$ by a discrete subgroup of isometries $\Gamma$, we obtain a new background which hosts a theory containing only the fields and operators invariant under $\Gamma$. As for the dual CFT$_{d}$, the same thing happens; the new theory after quotienting still has local conformal symmetry, but its states must also respect $\Gamma$. Furthermore, any fixed point singularities which cannot be avoided in any fundamental domain representation\footnote{This is as opposed to quotients whose fixed points \textit{can} be avoided by \textit{some} fundamental domain. For example, when quotienting AdS$_3$ by dilatation, we have a single fixed point in Poincar\'e coordinates, but we can avoid it to see that dilatation produces the locally AdS$_3$ two-sided BTZ\cite{Banados:1992gq}. The resulting state lives in a field theory without defects.} will introduce boundary points in the quotient space, turning the surviving CFT$_d$ states into BCFT$_d$ states.

In the case of AdS$_3$/CFT$_2$, even with such singularities, duality between the new theories on the orbifolded spaces should still be preserved for two reasons. First, each field or operator which is kept when quotienting the bulk should correspond with a field or operators which is kept when quotienting the conformal boundary. Secondly, for three bulk dimensions, the Einstein equations of motion imply that there are no bulk degrees of freedom, so the full spectrum of (B)CFT$_2$ states are represented by locally empty AdS$_3$ geometries.

As a result, we have the \textit{quotient-AdS$_3$/BCFT$_2$ correspondence}; when unavoidable fixed point singularities which extend to the conformal boundary enter the picture, well-defined quantum gravity defined on a particular quotient space of AdS$_3$ is dual to a BCFT$_2$ defined on the relevant space with boundary. The bulk is still locally AdS$_3$ away from any fixed points, since we only consider quotienting by isometries. Furthermore, we conjecture that the branes obtained in this fashion are always tensionless, since, loosely, quotienting does not affect the underlying theory.

How these holographic bulk spaces in correspondence with BCFTs actually look is not too different from the picture presented by Takayanagi in \cite{Takayanagi:2011zk}. Using his notation, if we start with a BCFT defined on a $d$-dimensional manifold $\mathcal{M}$, we can extend it to a holographic bulk by ``attaching" $\mathcal{M}$ to an asymptotically AdS spacetime $\mathcal{N}$ which is $(d+1)$-dimensional. $\mathcal{N}$ itself has a boundary,
\begin{equation}
\partial\mathcal{N} = \mathcal{Q} \cup \mathcal{M}.
\end{equation}

Furthermore, $\mathcal{Q}$ and $\mathcal{M}$ are themselves connected at \textit{their} boundaries,
\begin{equation}
\partial\mathcal{Q} = \partial\mathcal{M}.
\end{equation}

This is indeed what we see in Figure \ref{figs:parity2}, when we quotient by parity. The boundary of the bulk consists of the conformal boundary (dark blue), on which the BCFT is defined, and the tensionless brane (red region), which is anchored to the conformal boundary. We also see Takayanagi's construction realized in Figure \ref{figs:hyperSurf}, when quotienting by parity and inversion together.

In a sense, quotient-AdS$_3$/BCFT$_2$ provides a natural method of constructing AdS/BCFT configurations. We have explored this idea specifically in three bulk dimensions (quotient-AdS$_3$/BCFT$_2$), even arguing for duality between the \textit{theories},\footnote{The existence of dual states need not imply duality between theories; to prove duality between theories, we would need to confirm that \textit{all} BCFT states can be realized holographically by quotienting asymptotically AdS spaces, but the equations of motion will not be vacuum equations outside of three dimensions.} but, if quotienting by isometries, we should be able to explore particular, potentially dynamical BCFT states in higher dimensions through this procedure.

\subsection{Quotients and AdS/dCFT}\label{dCFT}

A \textit{defect CFT} or \textit{dCFT} has the local symmetries of a CFT, but is defined on a space with defects that have a particular codimension. BCFTs are a particular case of dCFTs, in which the defects are on a surface of codimension 1 (aptly called the boundary).

The properties of holographic dCFTs have been explored thoroughly throughout the literature\cite{Bachas:2001vj,DeWolfe:2001pq,Aharony:2003qf,Jensen:2013lxa,Linardopoulos:2020jck}. The key point is that, just as there is an AdS/BCFT correspondence, there is a much broader AdS/dCFT correspondence, in which the gravitational theory will also contain a defect\cite{Linardopoulos:2020jck}.

In light of the above work, this begs the question: is there a quotient formulation of AdS/dCFT? In AdS$_3$, as we have already explored the case of defects on a subspace of codimension 1, but we could also try to explore defects with different codimension.

For the codimension 2 case, one possibility is for the fixed points to make-up a line in AdS$_3$. Furthermore, for the dual field theory to also have defects, this line would need to reach the conformal boundary. One way to construct such a defect would be to use rotation of AdS$_3$ in Poincar\'e coordinates about the line $t = x = 0$ by an angle $\theta$, i.e. to quotient by a boost. The resulting dCFT$_2$ would have a conical defect, with deficit angle $2\pi - \theta$, at the origin.

More generically, for codimension $2$ bulk conical defects, \cite{Martinec:2001cf} discusses how quotient spaces using isometries described by $\mathbb{Z}_n$ transformations can host quantum gravity theories, and \cite{Lunin:2002fw,Arefeva:2016wek} even discuss the holographic nature of such orbifolds. Specifically, \cite{Arefeva:2016wek} describes duality between a theory on an AdS$_3$/$\mathbb{Z}_n$ orbifold and a CFT$_2$ on a cylinder whose geometry is controlled by the deficit angle. Indeed, the dual field theories for such bulk theories need not be dCFTs, at least in the context of quotients. This is because we may consider isometries which produce fixed points \textit{only} in the bulk, resulting in bulk defects but not conformal boundary defects.

For more general spacetimes in a larger number of dimensions $d$, one could consider exploring defects of specific codimension. The more natural cases of interest may either be codimension $1$ defects (boundaries) or codimension $d-1$ defects (specifically, lines consisting of conical singularities).

As dCFTs are a generalization of BCFTs, exploring how their states may be realized via quotient spaces constitutes a possible future direction of interest.

\section{Conclusions}\label{conclusion}

To summarize, we have used quotient spaces of AdS$_3$ in order to arrive at a natural construction of holographic duals for BCFT$_2$ states. In three bulk dimensions, this hints at duality between quantum gravity on quotients of AdS$_3$ and particular BCFT$_2$ configurations, both of which would have matching boundaries. Our results line-up with what is known regarding AdS/BCFT, but start with the AdS/CFT correspondence and use quotients in a more constructive manner. The true power of quotienting, as shown in the AdS$_3$ strip construction of Section \ref{parityInvQ}, lies in its ability to yield dynamical configurations with strong analytic control. Our approach could perhaps be used to explore the structure of more exotic BCFT and dCFT states.

There are several avenues available for future work. In no particular order, they are as follows.

\begin{itemize}
\item The quotient explored in Section \ref{parityInvQ} is specifically a BCFT state whose boundary is time-dependent. Thus, one could, in principle, compute quantities like entropy and complexity in order to see how they may evolve in time. Furthermore, one could attempt to understand such quantities holographically, in the bulk quotient space. Similar dynamical analysis could be performed for the constructions of \cite{Loran:2010qn,Loran:2010zy} involving spacelike branes, which can be interpreted as delta-function sources.\\

\item Is there a way to quotient AdS$_3$ which would produce multipartite BCFT$_2$ states? For a bipartite state, one possibility is to quotient by dilatation controlled by a scale factor $\lambda$, as is done to construct a static two-sided BTZ, then quotient again by a composition of inversion and parity (Section \ref{parityInvQ}). By choosing the factor $a$ in \eqref{inversionP} such that, 
\begin{equation}
a = \frac{\lambda}{2},
\end{equation}
the fixed points on the $t = 0$ slice would essentially cut the corresponding, underlying Riemann surface of the static two-sided BTZ in half. Thus, conceptually, this quotient may be a ``folding" of the BTZ, consequently producing a so-called \textit{BTZ half-space}. If this is true, then we could also probe entanglement in multipartite BCFT states.\\

\item The isometries in Section \ref{adsQuotient} have fixed points constituting codimension 1 subspaces. One could quotient in such a way that the fixed points live in subspaces of different codimension. This should lead to holographic descriptions of particular dCFT$_2$ or CFT$_2$ states, depending on where the fixed points lie.\\

\item We only explored two different isometries in this work. It would be interesting to see what other types of boundaries/branes can be obtained. A possibly lofty goal would be a full classification of the holographic BCFTs and dCFTs obtained by quotienting AdS$_3$, which could be realized by classifying the different types of defects that could be obtained via different isometries. \cite{Loran:2010zy} provides a partial classification of geometries featuring axisymmetry and spacelike/null branes. For timelike branes realized via a quotient, the tension may vanish.\\

\item We can explicitly ask what may happen when we quotient higher-dimensional AdS$_d$, with $d > 3$. Conceptually, we expect some version of quotient-AdS/BCFT to work for any dimension, but, as there are bulk degrees of freedom in higher-dimensional Einstein-Hilbert gravity, only considering a quotient of empty AdS$_d$ as a background may not provide bulk duals for every BCFT$_{d-1}$ state. In other words, while we may have dual \textit{states}, we may not have dual \textit{theories}. One may need to quotient spacetimes which are only asymptotically AdS$_d$, as well, but such spacetimes are not necessarily maximally symmetric, limiting the available number of independent isometries. Exploring this question could allow us to study higher-dimensional BCFT via quotienting.
\end{itemize}

\section{Acknowledgments}

I thank Elena C\'aceres and Muhammad A. Shehper for helpful conversations which inspired the direction of this work, as well as for reviewing the text. Additionally, I thank Andreas Karch for also looking over the manuscript and providing feedback, both regarding my calculations in Section \ref{adsQuotient} and concerning the broader story of holographic defect theories.\\

I am supported by the National Science Foundation (NSF) Grants No. PHY-1820712 and No. PHY-1620610.

\bibliographystyle{jhep}
\bibliography{multi}

\providecommand{\href}[2]{#2}\begingroup\raggedright\begin{thebibliography}{10}

\bibitem{Brown:1986nw}
J.~D. Brown and M.~Henneaux, \emph{{Central Charges in the Canonical
  Realization of Asymptotic Symmetries: An Example from Three-Dimensional
  Gravity}}, \href{https://doi.org/10.1007/BF01211590}{\emph{Commun. Math.
  Phys.} {\bfseries 104} (1986) 207}.

\bibitem{Banados:1992gq}
M.~Banados, M.~Henneaux, C.~Teitelboim and J.~Zanelli, \emph{{Geometry of the
  (2+1) black hole}}, \href{https://doi.org/10.1103/PhysRevD.48.1506,
  10.1103/PhysRevD.88.069902}{\emph{Phys. Rev.} {\bfseries D48} (1993) 1506}
  [\href{https://arxiv.org/abs/gr-qc/9302012}{{\ttfamily gr-qc/9302012}}].

\bibitem{Skenderis:2009ju}
K.~Skenderis and B.~C. van Rees, \emph{{Holography and wormholes in 2+1
  dimensions}}, \href{https://doi.org/10.1007/s00220-010-1163-z}{\emph{Commun.
  Math. Phys.} {\bfseries 301} (2011) 583}
  [\href{https://arxiv.org/abs/0912.2090}{{\ttfamily 0912.2090}}].

\bibitem{Balasubramanian:2014hda}
V.~Balasubramanian, P.~Hayden, A.~Maloney, D.~Marolf and S.~F. Ross,
  \emph{{Multiboundary Wormholes and Holographic Entanglement}},
  \href{https://doi.org/10.1088/0264-9381/31/18/185015}{\emph{Class. Quant.
  Grav.} {\bfseries 31} (2014) 185015}
  [\href{https://arxiv.org/abs/1406.2663}{{\ttfamily 1406.2663}}].

\bibitem{Caceres:2019giy}
E.~Caceres, A.~Kundu, A.~K. Patra and S.~Shashi, \emph{{A Killing Vector
  Treatment of Multiboundary Wormholes}},
  \href{https://doi.org/10.1007/JHEP02(2020)149}{\emph{JHEP} {\bfseries 02}
  (2020) 149} [\href{https://arxiv.org/abs/1912.08793}{{\ttfamily
  1912.08793}}].

\bibitem{Martinec:2001cf}
E.~J. Martinec and W.~McElgin, \emph{{String theory on AdS orbifolds}},
  \href{https://doi.org/10.1088/1126-6708/2002/04/029}{\emph{JHEP} {\bfseries
  04} (2002) 029} [\href{https://arxiv.org/abs/hep-th/0106171}{{\ttfamily
  hep-th/0106171}}].

\bibitem{Horowitz:2001uh}
G.~T. Horowitz and T.~Jacobson, \emph{{Note on gauge theories on M / Gamma and
  the AdS / CFT correspondence}},
  \href{https://doi.org/10.1088/1126-6708/2002/01/013}{\emph{JHEP} {\bfseries
  01} (2002) 013} [\href{https://arxiv.org/abs/hep-th/0112131}{{\ttfamily
  hep-th/0112131}}].

\bibitem{Cardy:2004hm}
J.~L. Cardy, \emph{{Boundary conformal field theory}},
  \href{https://arxiv.org/abs/hep-th/0411189}{{\ttfamily hep-th/0411189}}.

\bibitem{Karch:2000gx}
A.~Karch and L.~Randall, \emph{{Open and closed string interpretation of SUSY
  CFT's on branes with boundaries}},
  \href{https://doi.org/10.1088/1126-6708/2001/06/063}{\emph{JHEP} {\bfseries
  06} (2001) 063} [\href{https://arxiv.org/abs/hep-th/0105132}{{\ttfamily
  hep-th/0105132}}].

\bibitem{Takayanagi:2011zk}
T.~Takayanagi, \emph{{Holographic Dual of BCFT}},
  \href{https://doi.org/10.1103/PhysRevLett.107.101602}{\emph{Phys. Rev. Lett.}
  {\bfseries 107} (2011) 101602}
  [\href{https://arxiv.org/abs/1105.5165}{{\ttfamily 1105.5165}}].

\bibitem{Fujita:2011fp}
M.~Fujita, T.~Takayanagi and E.~Tonni, \emph{{Aspects of AdS/BCFT}},
  \href{https://doi.org/10.1007/JHEP11(2011)043}{\emph{JHEP} {\bfseries 11}
  (2011) 043} [\href{https://arxiv.org/abs/1108.5152}{{\ttfamily 1108.5152}}].

\bibitem{Alishahiha:2011rg}
M.~Alishahiha and R.~Fareghbal, \emph{{Boundary CFT from Holography}},
  \href{https://doi.org/10.1103/PhysRevD.84.106002}{\emph{Phys. Rev.}
  {\bfseries D84} (2011) 106002}
  [\href{https://arxiv.org/abs/1108.5607}{{\ttfamily 1108.5607}}].

\bibitem{Cavalcanti:2018pta}
A.~G. Cavalcanti, D.~Melnikov and M.~R. Silva, \emph{{Studies of Boundary
  Entropy in AdS/BCFT}},  \href{https://arxiv.org/abs/1808.07966}{{\ttfamily
  1808.07966}}.

\bibitem{Sato:2020upl}
Y.~Sato, \emph{{Boundary entropy under ambient RG flow in the AdS/BCFT model}},
   \href{https://arxiv.org/abs/2004.04929}{{\ttfamily 2004.04929}}.

\bibitem{Sully:2020pza}
J.~Sully, M.~Van~Raamsdonk and D.~Wakeham, \emph{{BCFT entanglement entropy at
  large central charge and the black hole interior}},
  \href{https://arxiv.org/abs/2004.13088}{{\ttfamily 2004.13088}}.

\bibitem{Bachas:2001vj}
C.~Bachas, J.~de~Boer, R.~Dijkgraaf and H.~Ooguri, \emph{{Permeable conformal
  walls and holography}},
  \href{https://doi.org/10.1088/1126-6708/2002/06/027}{\emph{JHEP} {\bfseries
  06} (2002) 027} [\href{https://arxiv.org/abs/hep-th/0111210}{{\ttfamily
  hep-th/0111210}}].

\bibitem{Loran:2010qn}
F.~Loran and M.~Sheikh-Jabbari, \emph{{O-BTZ: Orientifolded BTZ Black Hole}},
  \href{https://doi.org/10.1016/j.physletb.2010.08.022}{\emph{Phys. Lett. B}
  {\bfseries 693} (2010) 184}
  [\href{https://arxiv.org/abs/1003.4089}{{\ttfamily 1003.4089}}].

\bibitem{Loran:2010zy}
F.~Loran and M.~Sheikh-Jabbari, \emph{{Orientifolded Locally $AdS_3$
  Geometries}},
  \href{https://doi.org/10.1088/0264-9381/28/2/025013}{\emph{Class. Quant.
  Grav.} {\bfseries 28} (2011) 025013}
  [\href{https://arxiv.org/abs/1008.0462}{{\ttfamily 1008.0462}}].

\bibitem{Jensen:2013lxa}
K.~Jensen and A.~O'Bannon, \emph{{Holography, Entanglement Entropy, and
  Conformal Field Theories with Boundaries or Defects}},
  \href{https://doi.org/10.1103/PhysRevD.88.106006}{\emph{Phys. Rev. D}
  {\bfseries 88} (2013) 106006}
  [\href{https://arxiv.org/abs/1309.4523}{{\ttfamily 1309.4523}}].

\bibitem{Randall:1999vf}
L.~Randall and R.~Sundrum, \emph{{An Alternative to compactification}},
  \href{https://doi.org/10.1103/PhysRevLett.83.4690}{\emph{Phys. Rev. Lett.}
  {\bfseries 83} (1999) 4690}
  [\href{https://arxiv.org/abs/hep-th/9906064}{{\ttfamily hep-th/9906064}}].

\bibitem{DeWolfe:2001pq}
O.~DeWolfe, D.~Z. Freedman and H.~Ooguri, \emph{{Holography and defect
  conformal field theories}},
  \href{https://doi.org/10.1103/PhysRevD.66.025009}{\emph{Phys. Rev. D}
  {\bfseries 66} (2002) 025009}
  [\href{https://arxiv.org/abs/hep-th/0111135}{{\ttfamily hep-th/0111135}}].

\bibitem{Aharony:2003qf}
O.~Aharony, O.~DeWolfe, D.~Z. Freedman and A.~Karch, \emph{{Defect conformal
  field theory and locally localized gravity}},
  \href{https://doi.org/10.1088/1126-6708/2003/07/030}{\emph{JHEP} {\bfseries
  07} (2003) 030} [\href{https://arxiv.org/abs/hep-th/0303249}{{\ttfamily
  hep-th/0303249}}].

\bibitem{Linardopoulos:2020jck}
G.~Linardopoulos, \emph{{Solving holographic defects}}, {\emph{PoS} {\bfseries
  CORFU2019} (2019) 141} [\href{https://arxiv.org/abs/2005.02117}{{\ttfamily
  2005.02117}}].

\bibitem{Lunin:2002fw}
O.~Lunin and S.~D. Mathur, \emph{{Rotating deformations of AdS(3) x S**3, the
  orbifold CFT and strings in the pp wave limit}},
  \href{https://doi.org/10.1016/S0550-3213(02)00677-6}{\emph{Nucl. Phys. B}
  {\bfseries 642} (2002) 91}
  [\href{https://arxiv.org/abs/hep-th/0206107}{{\ttfamily hep-th/0206107}}].

\bibitem{Arefeva:2016wek}
I.~Y. Aref'eva and M.~A. Khramtsov, \emph{{AdS/CFT prescription for
  angle-deficit space and winding geodesics}},
  \href{https://doi.org/10.1007/JHEP04(2016)121}{\emph{JHEP} {\bfseries 04}
  (2016) 121} [\href{https://arxiv.org/abs/1601.02008}{{\ttfamily
  1601.02008}}].

\end{thebibliography}\endgroup
\end{document}